\documentclass{kluwer}
\usepackage{epsfig}

\def\be{\begin{equation}}
\def\ee{\end{equation}}

\def\ga{{\leavevmode\kern0.3em\raise.3ex\hbox{$>$}
\kern-0.8em\lower.7ex \hbox{$\sim$}\kern0.3em}}
\def\la{{\leavevmode\kern0.3em\raise.3ex\hbox{$<$}
\kern-0.8em\lower.7ex \hbox{$\sim$}\kern0.3em}}

\begin{document}
\begin{article}
\begin{opening}

\runningtitle{Solar cycle variations of large scale flows in the Sun}
\runningauthor{S. Basu and H. M. Antia}

\title{SOLAR CYCLE VARIATIONS OF LARGE SCALE FLOWS IN THE SUN}

\author{Sarbani \surname{Basu}}
\institute{Institute for Advanced Study,
Olden Lane, Princeton NJ 08540, U. S. A. and
Astronomy Department, Yale University, P.O. Box 208101 New Haven,
CT 06520-8101 USA}
\author{H. M. \surname{Antia}}
\institute{Tata Institute of Fundamental Research,
Homi Bhabha Road, Mumbai 400005, India}
\date{\today}

\begin{abstract}
Using data from the Michelson Doppler Imager (MDI) instrument
on board the Solar and Heliospheric Observatory (SOHO), we study the
large-scale velocity fields in the outer part of the solar convection zone
using the ring diagram technique. We use observations from four
different times to study possible temporal variations in flow velocity.
We find definite changes in both the zonal and meridional
components of the flows. The amplitude of the zonal flow appears to
increase with solar activity and the flow pattern also shifts towards
lower latitude with time. 
\end{abstract}

\end{opening}

\section{Introduction}
Ring diagram analysis has been extensively used to infer horizontal
flows in the outer part of the solar convection zone (Hill 1988;
Patr\'on et al.~1997; Basu, Antia and Tripathy 1999).
This technique is based on the study of three-dimensional
power spectra of solar
p-modes on a part of the solar surface, and can be used
to study the variation in flow velocity with latitude, longitude
and time. The latitudinal variation of flow velocities in both
the rotational and meridional component has been
extensively studied (Schou and Bogart 1998; Basu et al.~1999; 
Gonzalez Hernandez et al.~1999)
and there is a reasonable agreement between results obtained
by different workers. There is some indication that the flow
velocity also changes with longitude and time (Patr\'on et al.~1998),
but it is not
clear if some of these variations are due to uncertainties in
estimating the velocities or some local influence like the presence of
active regions.

Since the large scale flows, like the differential rotation and
meridional flows are expected to play an important role in
the functioning of the solar dynamo (Choudhuri, Schussler and Dikpati~1995;
Brummell, Hurlburt and Toomre~1998), one may expect some changes in
these flow patterns over the solar cycle. These variations may give us
some information about how the solar dynamo operates.
From the splittings of global f-modes Schou~(1999) has found
variations  in the zonal flow pattern, which are similar to the torsional
oscillations observed at the solar surface. Thus it appears that
these oscillations penetrate into the deeper layers.
It would thus be interesting to study
possible variation in these large scale flows with solar activity
and how far deep these flows extend. 
With the high quality data collected by Michelson Doppler Imager (MDI)
on board SOHO, over the last 3 years it has been possible to study
these temporal variations and
in this work we attempt to study possible temporal variation in
large scale flows.
For simplicity, we have only considered the longitudinal averages,
which contain information about the latitudinal variation
in these flows. For this purpose, at each latitude
we have summed the spectra obtained for different longitudes to get
an average spectrum which has information on the average flow velocity
at each latitude. The averaging helps us in improving the reliability
of results thus allowing us to infer relatively small variations in the
flow velocities.
We have studied both the rotational and meridional components
of the flow velocity.

\section{Technique}

In this work we use the data obtained by the
MDI instrument to measure the flow velocities at different epochs.
We use four sets of data. Set 1 is data 
 for May/June 1996, Set 2 is 
 for July 1996, Set 3 is  for April 1997
and Set 4 is for January/February 1998.
These data were taken at various phases of the current solar
cycle.   
The data consist of three dimensional power spectra obtained
from full disc Dopplergrams. The Dopplergrams were taken at a
cadence of 1 minute. The area being studied was tracked at the
surface rotation rate. To minimize the effect of foreshortening
we have only used data for the central 
meridian.  Each power spectrum was obtained from
a time series of 1664 images covering $15^\circ$ in longitude and
latitude. Successive spectra are
separated by $15^\circ$ in heliographic longitude of the central
meridian. For each longitude, we have used 15 spectra centered at
latitudes ranging from $52.5^\circ$N to $52.5^\circ$S with a spacing
of $7.5^\circ$ in latitude.

Set 1 consists of spectra from twelve longitudes, starting
from central meridian at $105^\circ$
to $15^\circ$ of Carrington Rotation 1909 and 
$360 ^\circ$  to $300^\circ$ of rotation 1910.
Set 2 consists of twelve longitudes from Carrington rotation 1911,
central meridian at $285^\circ$ to $120^\circ$,
set 3 contains eight longitudes from Carrington rotation 1921, central meridian at
$120^\circ$ to $15^\circ$ and
set 4 contains nine longitudes from Carrington rotation 1932, central meridian at
$360^\circ$ to $240^\circ$.
The choice of these data sets was dictated by the availability of
data at the time this analysis was carried out.

To extract the flow velocities and other mode parameters from the
three dimensional power spectra we fit a model with asymmetric peak profiles
(Basu and Antia 1999) specified by:
\begin{eqnarray}
P(k_x,k_y,\nu)&=
{\exp(A_0+(k-k_0)A_1+A_2({k_x\over k})^2+
A_3{k_xk_y\over k^2})(S^2+(1+Sx)^2)\over
x^2+1}\nonumber \\
\noalign{\medskip}
& \qquad +{e^{B_1}\over k^3}+ {e^{B_2}\over k^4},
\label{eq:model}
\end{eqnarray}
where
\be
x={\nu-ck^p-U_xk_x-U_yk_y\over w_0+w_1(k-k_0)},
\label{eq:xx}
\ee
$k^2=k_x^2+k_y^2$, $k$ being the total wave number,
and the 13 parameters $A_0, A_1, A_2, A_3, c, p,
U_x, U_y, w_0, w_1, S, B_1$ and $B_2$ are determined by fitting the spectra
using a maximum likelihood approach (Anderson, Duvall and Jefferies~1990).
Here, $k_0$ is the central value of $k$ in the fitting interval and
$\exp(A_0)$ is the mean power in the ring.
The coefficient
$A_1$ accounts for the variation in power with $k$ in the fitting interval,
while $A_2$ and $A_3$ terms account for the variation of power along the
ring. The term $ck^p$ is the mean frequency, while
$U_xk_x$ and $U_yk_y$ represent the shift in
frequency due to large scale flows and the fitted values of $U_x$ and
$U_y$ give the average flow velocity over the region covered by the
power spectrum and the depth range where the corresponding mode is trapped.
The mean half-width is given by $w_0$, while $w_1$ takes care of
the variation in half-width with $k$ in the fitting interval.
The terms involving $B_1,B_2$ define the background power, which
is assumed to be of the same form as that used by Patr\'on et al.~(1997).
$S$ is a parameter that controls the
asymmetry, and the form of asymmetry is the same as that
prescribed by Nigam and Kosovichev~(1998).
The details of how the fits are obtained
can be found in Basu et al.~(1999) and Basu and Antia (1999).

The fitted $U_x$ and $U_y$ for each mode represents an average of the
velocities in the $x$ and $y$ directions
over the entire region in horizontal extent and over the vertical
region where the mode is trapped.
We can invert the fitted $U_x$ (or $U_y$) for a set of modes
to infer the variation in
horizontal flow velocity $u_x$ (or $u_y$) with depth.
We use the Regularized Least
Squares (RLS) as well as  the Optimally Localized Averages (OLA)
techniques for inversion as outlined by Basu et al.~(1999).
The results obtained
by these two independent inversion techniques are compared to test the
reliability of inversion results.

\begin{figure}
\centerline{\epsfig{file=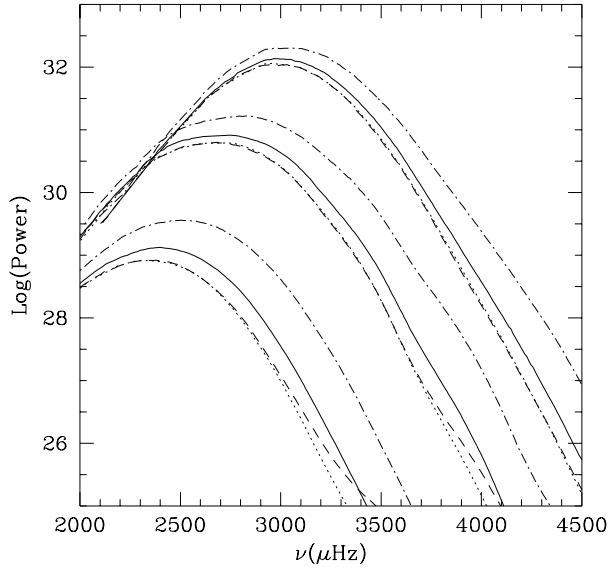,width=8.0 true cm}}
\caption{
The power in ridges for $n=0,1,2$ at different times is plotted against
frequency, for region centered at the equator.
The power is in arbitrary units. The continuous line is for
Set 1, the dotted line for Set 2, the dashed line for Set 3
and dot-dashed line for Set 4.
}
\end{figure}

\section{Results}
We have fitted the four sets of summed spectra to calculate various
mode parameters.
In this work we are mainly concerned with the variation in flow velocities
with time, but it is well known that other quantities also vary with
time. The variation of mean frequency with time is well-known, but it
is difficult to determine this small change using ring diagram technique
as the image scale is known to have changed with time to some extent.
We do not find any significant variation in the width but the power
in modes shows some temporal variation.
Fig.~1 shows the power in $n=0,1,2$ ridges in equatorial region
as a function of frequency for each of the four sets of spectra.
It is clear that power in Set 4, which covers a region of high activity
is larger as compared to the other three sets which cover period of low
activity. If this variation is real then the power in modes appears to
increase with activity and the increase is quite significant at higher
frequencies. Similar trend is seen in spectra centered at different
latitudes. The high frequency modes which penetrate into the near surface
region are more likely to be influenced by surface activity and thus
a larger increase in their amplitude indicates that the driving mechanism
operates in the near surface region, where the magnetic field
is known to change with activity.

\begin{figure}
\centerline{\epsfig{file=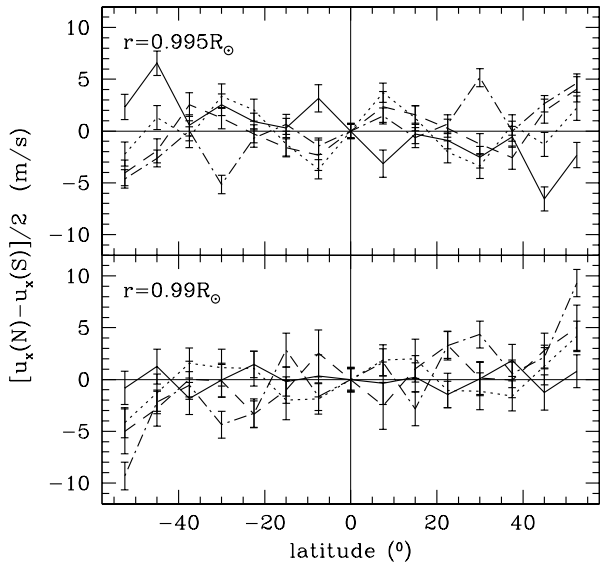,width=8.0 true cm}}
\caption{
The north-south antisymmetric part of the rotation velocity,
i.e., $[u_x(\hbox{N})-u_x(\hbox{S})]/2$,
plotted as a function of latitude at two different depths for all 4 sets
of data.
The line-types are the same as those in Fig.~1.
}
\end{figure}

\begin{figure}
\centerline{\epsfig{file=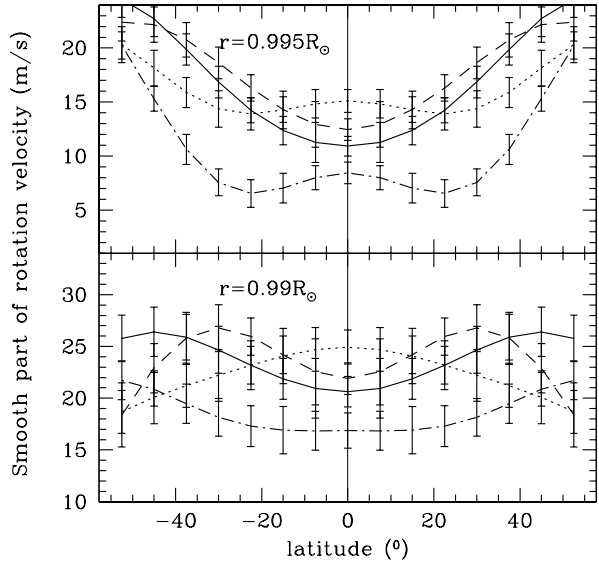,width=8.0 true cm}}
\caption{
The smooth part of the solar rotation velocity (after subtraction of the 
rate at which the regions on the Sun were tracked)
plotted as a function of latitude at two different depths for all 4 sets
of data.
The line-types are the same as those in Fig.~1.
}
\end{figure}

\subsection{The rotation velocity}
The horizontal velocity $u_x$ is known to be dominated by the 
rotation rate, since the area under observation is tracked only
at the surface rotation rate, and the solar rotation rate is known to
vary with depth. Thus the measured $u_x$ arises from the difference
in rotation rate and the tracking rate.
The ring diagram technique allows us to determine
the north-south antisymmetric component of rotation rate
which cannot be inferred by the splittings of global modes.
Fig.~2 shows the results at two different depths. There appears to be
some temporal variation in this component, which is comparable to the
error estimates. There is no obvious pattern in temporal variation
and it is not clear if the apparent difference between different sets
is not due to some systematic errors. Of course, it is
possible that the antisymmetric
component of the rotation velocity  varies  on time scales of the
order of the rotation period, which is the separation between Sets 1 and 2.
Such variations in the rotation rate at solar surface have been seen in
Doppler measurements (Hathaway et al.~1996; Ulrich~1998).

Following  Kosovichev and Schou (1997), it is possible to 
decompose $u_x$ into two components, a smooth part (expressed in
terms of  $\cos\theta$, $\cos^3\theta$ and $\cos^5\theta$, $\theta$
being the latitude) and the remaining part, which is identified
with the zonal flows.  
Fig.~3 shows the smooth part of the rotation rate for all 4 data sets.
This result automatically has the tracking-rate which is mean rotation
velocity at solar surface, subtracted. There is some variation
in the rotation velocity with time. However,
the change between the low activity sets (1 to 3) may not be
statistically significant.

\begin{figure}
\centerline{\epsfig{file=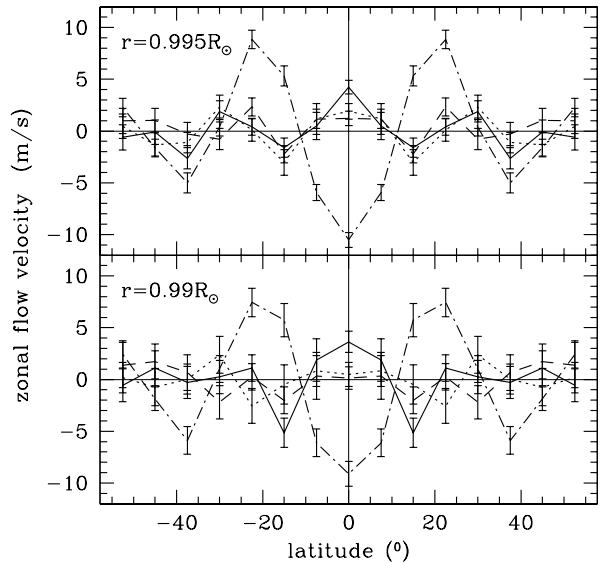,width=8.0 true cm}}
\caption{
The zonal flow velocity plotted as a function of latitude at two different
depths. 
The line-types are the same as those in Fig.~1.
}
\end{figure}

\begin{figure}
\centerline{{\epsfig{file=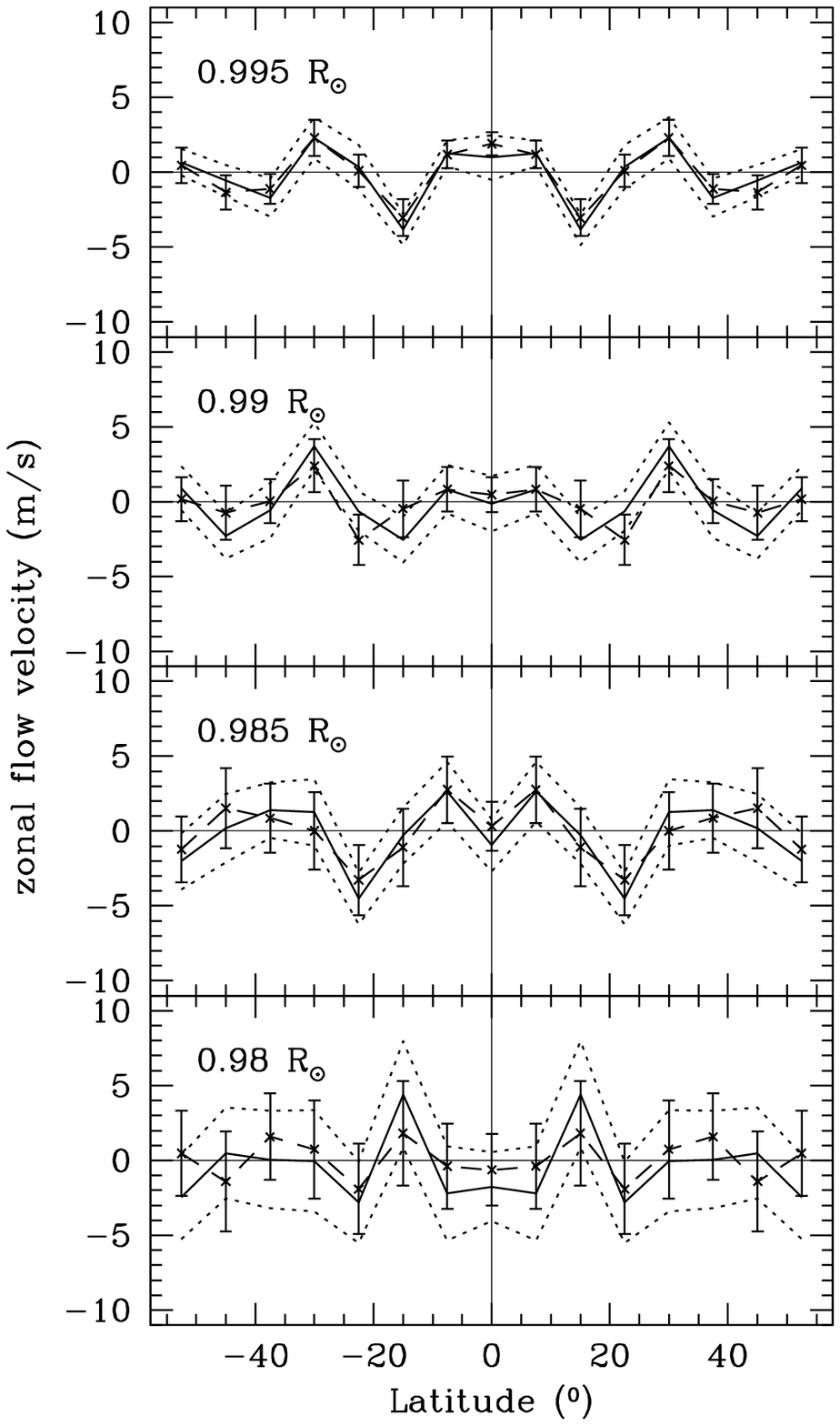,width=5.5 true cm}}\hfill
{\epsfig{file=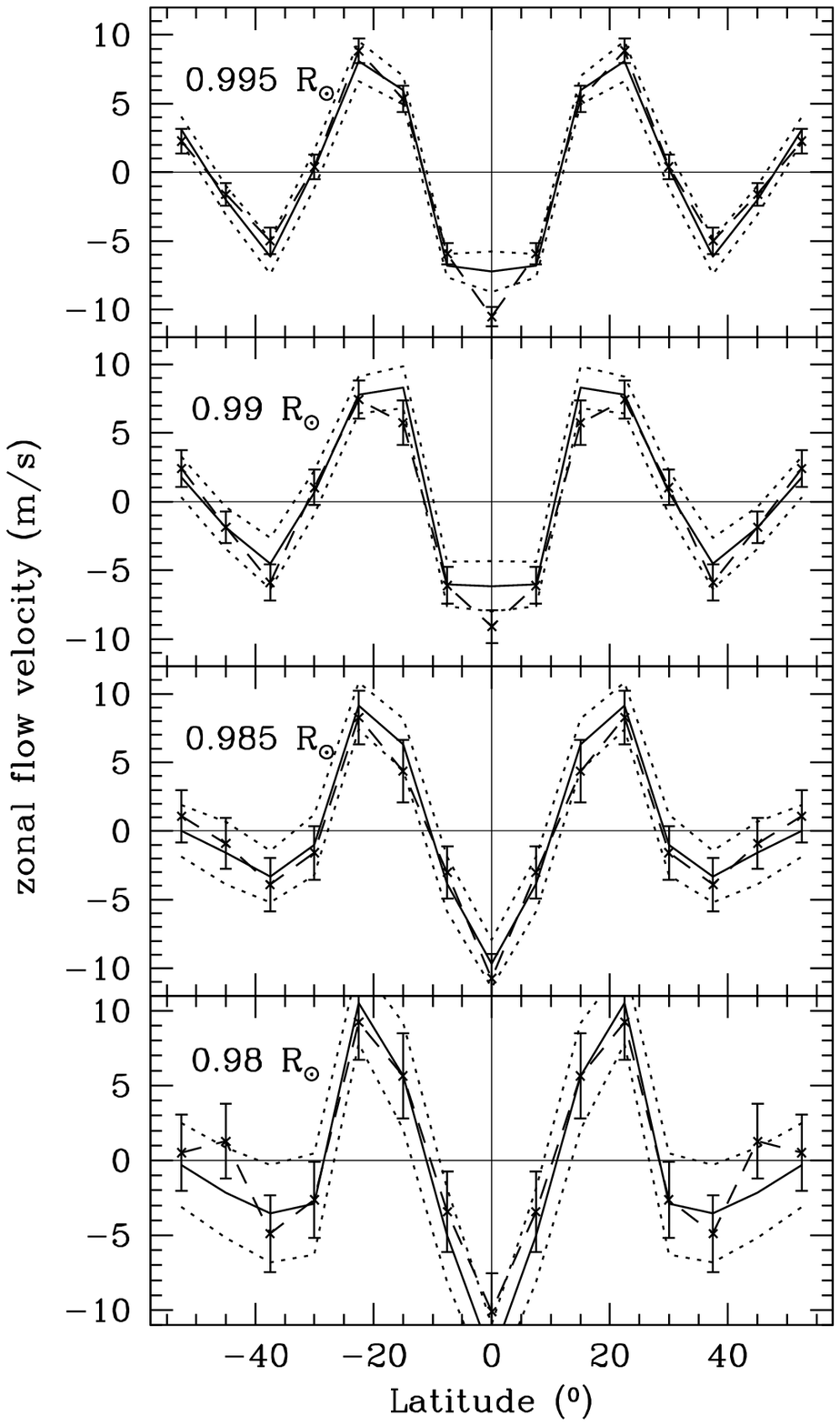,width=5.5 true cm}}}
\caption{The zonal flow velocity plotted as a function of latitude at four
different depths as marked in the figure.
The panels on the left shows the result for
Set 2 (low activity), while the panel on the right show the results
for set 4 (high activity). In each panel, the points joined by the
dashed lines are the results of an optimally localized averages
(OLA) inversion. The continuous lines are the results of a regularized
least squares (RLS) inversion with dotted lines showing the $1\sigma$
error limits.
}
\end{figure}

Results from global f-mode splittings show that 
there is a variation in the  zonal flow pattern with time
(Schou 1999) ---  the position of maximum and minimum zonal flow velocities
drift equator-wards similar to the torsional oscillation pattern
seen on the solar surface.
Similar results have been obtained by Howe, Komm and Hill (2000)
and Toomre et al.~(2000) from inversion of frequency splittings.
The ring diagram analysis provides a better depth resolution in
near surface region as compared to the global modes and hence it can
be employed to study the variation with depth in zonal flow pattern.
However, it is not clear if the non-smooth
part of the rotation velocity is the same as the torsional oscillations,
since the latter is the time variation in rotation velocity, and there
could be some variation in smooth component  which would also contribute
to the torsional oscillations. This variation is in fact seen from
our results in Fig.~3. Ideally, we can determine an average
rotation rate over a long time period and subtract it from each
measurement to get the time variation, which can possibly be
identified with torsional oscillations.  With only four measurements
in time it is not possible to take any meaningful average and hence
we attempt to identify the zonal flows with the non-smooth component
as has been done by Kosovichev and Schou~(1997) and Schou~(1999).

It is known from earlier studies that 
the symmetric component of the zonal flow velocities agree well with the
results from f-mode analysis (Basu et al.~1999).
Fig.~4 shows the symmetric component of the zonal flow velocities for all the
data sets at two different depths. Note that there is some variation in the
results with time. The change is particularly significant between sets
3 and 4. Solar activity  is known to have
changed quite rapidly between these two epochs. The latitudinal
resolution of our study
is not enough to clearly judge whether the changes between results
of Sets 1, 2 and 3 are statistically significant, however, there appears
to be a shift towards equator in the first maxima in each hemisphere
at a radial distance of $0.995R_\odot$.
Results from  Set 4  show the equatorial region to be  rotating slower 
than the smoothed
average rotation rate, while results from the the other sets show a faster rotation.
This feature is also seen in the results obtained from f-mode
splittings (Schou 1999).

To study the changes in the zonal flow pattern with depth we
concentrate on Sets 2 and 4. Fig.~5 shows the zonal flow velocities
for sets 2 and 4 at four different depths obtained using both RLS and OLA inversion
techniques. The results obtained by the two inversion techniques
agree with each other in all cases. 
We see that Set 4
in general has much larger flow velocities than Set 2.
Furthermore, for the low
activity set the direction of the  flow appears to change sign as
one goes deeper. This is not seen for the Set 4.  However, since the
overall flow velocity is small for Set 2, it is somewhat
difficult to judge the significance of the change.
From the results for Set 4, it appears that the zonal flow pattern
persists up to a depth of at least $0.02R_\odot$ during high activity
period.
Because of low amplitude it is not very
clear if the zonal flow pattern penetrates to the depth of $0.02R_\odot$
during the low activity period too.
There may be some change in pattern with depth and it is possible that
the reversal in sign of zonal flow velocity
at equator occurs after the solar minimum and starts at deeper layers,
gradually advancing towards the surface, i.e.,   the reversal occurs
earlier in deeper layers.
Surface observations appear to show that the zonal flow
patterns is not well defined during the time around the solar maximum
(Ulrich 1998).
Thus it would be interesting to check what happens in deeper layers
as we approach the maximum of cycle 23.
Fig.~6 shows the magnitude of zonal flow velocity averaged over all
latitudes included in this study, at a depth of $0.005R_\odot$,
plotted together with the
mean daily sunspot number. Although we have a very small sample,
it appears that the mean flow velocity increases with increase in 
solar activity.
Similar results can be obtained from study of global
modes, for which data is available for larger duration, spanning a
wider range of activity level.

\begin{figure}
\centerline{\epsfig{file=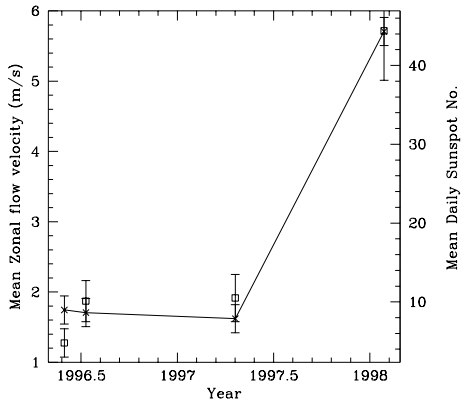,width=6.0 true cm}}
\caption{The average zonal flow velocity at $r=0.995R_\odot$
plotted as a function of time (continuous line with crosses).
The squares show the mean daily sunspot number during the same
period when the observations were made, with the scale on the right-hand
axis.
}
\end{figure}

\begin{figure}
\centerline{\epsfig{file=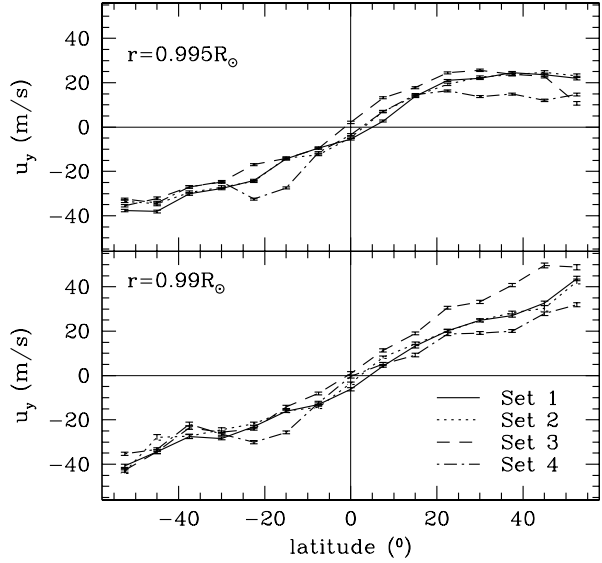,width=8.0 true cm}}
\caption{
The meridional flow velocities obtained from the 4 data sets
plotted as a function of latitude at two different depths marked
in the figure.
The line types are the same as those in Fig.~1.
}
\end{figure}

\subsection{The meridional flow}
The velocity component $u_y$ is the meridional
flow. Earlier  results have shown that the predominant flow
pattern at the surface is from the equator to the two poles
(Giles et al.~1997; Basu et al.~1999;  Braun and Fan 1998). 
Fig.~7 shows the meridional flow velocity
for the 4 sets at two different depths. It is clear that once again
the Set 4 pattern is significantly different than the others, and in particular the
amplitudes of high order component appears to be larger giving
rise to some oscillations over the smooth pattern.

To take a more detailed look at the time variation, following
Hathaway et al.~(1996) we try to
express the  meridional component as
\be
u_y(r,\phi)=-\sum_i a_i(r) P_i^1(\cos(\phi))
\label{eq:comp}
\ee
where $\phi$ is the colatitude, and $P_i^1(x)$ are associated Legendre
polynomials. The first six terms in this expansion are found to be
significant and their amplitudes are shown in Fig.~8.
The component $a_6$ seems to show the most significant changes in the
outer layers where the inversions are most reliable.
The amplitude of $a_6$ around the depth of $0.005R_\odot$ appears to
increase with solar activity and may be correlated to the
mean daily sunspot number.
The amplitude of the dominant component, $a_2$ appears to have
decreased slightly during the
high activity period at depths around $0.005R_\odot$.
The odd terms in the expansion, which represent the
north-south symmetric component of meridional flow, show much larger
variation, particularly the first term $a_1(r)$. The amplitude $a_1(r)$
of the first component appears to have reduced by about 5 m/s in Set 4
as compared to Sets 1--3.
In deeper layers the odd components
of meridional flow appears to vary anomalously for Set~3. 
It is not clear if these variations in the odd components
are due to instrumental effects arising from change in
pointing errors,  some systematic error in our analysis process,
or whether they represent real changes in the subsurface flows
with activity. Surface observations of meridional flows also reveal
short term variations in the various components (Hathaway et al.~1996)
which may correspond to some of the changes in the odd components.
More detailed study involving larger number of data sets in time would be 
required to understand these variations.

\begin{figure}
\centerline{\epsfig{file=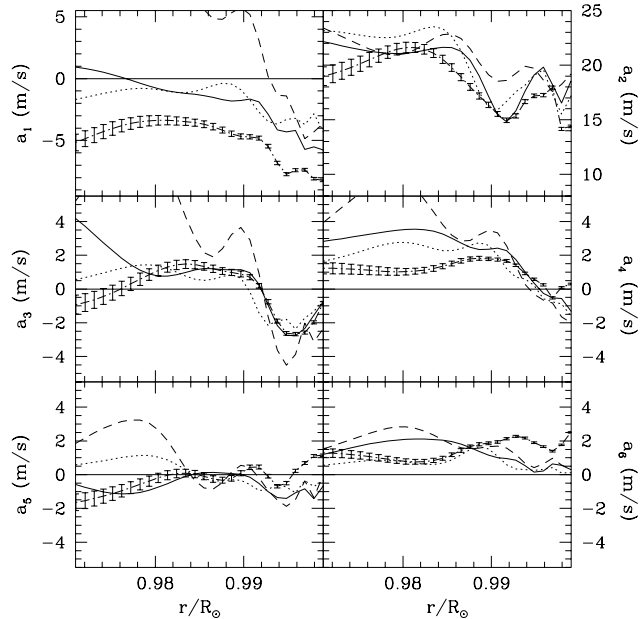,width=9.0 true cm}}
\caption{
The amplitude of different components of the meridional flows as defined in
Eq.~(\ref{eq:comp}). The line types are the same as those in Fig.~1.
Only one set of error-bars have been shown for the sake of clarity.
}
\end{figure}

\section{Conclusions}
We find that the large-scale flows in the upper part of the 
solar convection zone vary with time. We find that the
amplitude of the zonal flow increases with increase in solar 
activity and besides there is some shift in the zonal flow pattern with
time, which is similar to the torsional oscillations observed at the
solar surface.
The zonal flow pattern appears to persist up to a depth of at least
$0.02R_\odot$
during period of high activity, when the amplitude is large enough for
a reliable determination of the flow pattern. During the low activity period
it is not clear if the zonal flow pattern penetrates to these depths.
It may be noted that from the analysis of splittings of global modes
Howe et al.~(2000) and Toomre et al.~(2000) 
find that the zonal flow pattern persists to greater depths.
Further, it is found that during high activity period in 1998 the
equatorial region was rotating slower than the smoothed average
rotation rate. It was rotating faster at other times.
There is probably some depth dependence of the flow pattern during the low
activity period, but its significance is not clear because of the low
amplitude of the flows.

The meridional component of the flows also changes with time.
If the meridional flow
velocities are decomposed on a basis of associated Legendre
polynomials, the high order anti-symmetric component corresponding
to $P_6^1(\cos\phi)$ shows an increase in amplitude
with solar activity.  There are also significant changes in the symmetric
component of meridional flow, but it is not clear if this  is not an instrumental
effect or due to some systematic errors in analysis.

\begin{acknowledgements}
This work  utilizes data from the Solar Oscillations
Investigation / Michelson Doppler Imager (SOI/MDI) on the Solar
and Heliospheric Observatory (SOHO).  SOHO is a project of
international cooperation between ESA and NASA.
The authors would like to thank the SOI Science Support Center
and the SOI Ring Diagrams Team for assistance in data
processing. The data-processing modules used were
developed by Luiz A. Discher de Sa and Rick Bogart, with
contributions from Irene Gonz\'alez Hern\'andez and Peter Giles.
\end{acknowledgements}

\end{article}
\end{document}